\documentstyle[prl,twocolumn,aps,epsf]{revtex}

\begin{document}
 \tolerance 50000
\def\d{\dagger}
\draft

\twocolumn[\hsize\textwidth\columnwidth\hsize\csname@twocolumnfalse\endcsname

\title{ Low Energy Properties of Ferrimagnetic 2-leg Ladders: a Lanczos study}

\author{A.~Langari$^{\dagger}$  and M.A. Mart\'{\i}n-Delgado$^{\ast}$,
}
\address{
$^{\dagger}$ Max-Planck-Institut f\"ur Physik komplexer Systeme, N\"othnitzer Strasse 38,
D-01187 Dresden, Germany
\\
$^{\ast}$Departamento de F\'{\i}sica Te\'orica I, Universidad Complutense,
28040-Madrid, Spain.
}

\maketitle

\begin{abstract}
We apply the Lanczos method to a 2-leg ladder with mixed spins of
magnitudes $(S_1,S_2)=(1,1/2)$ located at alternating positions
along the ladder. The effect of dimerization $\gamma$ is also
considered according to two different patterns. A Spin Wave
Theory (SWT) is applied to this model predicting one gapless
branch with ferromagnetic properties and another gapful branch
with antiferromagnetic nature as low energy excitations of the
model. We compute the ground state energies, Ferro- and
AF-excitation gaps, magnetizations and correlation functions as a
function of $J'$ and $\gamma$ which results into a fine estimate
of the phase diagram. The Lanczos results are compared with the
SWT analysis and a qualitative agreement is found but with
numerical discrepancies. We also study numerically the
Spin-Peierls instability and find that it is conditional for any
value of $ J' \in (0, 2)$ and both  dimerization patterns.
\begin{center}
\parbox{14cm}{}

\end{center}
\end{abstract}

\vspace{-0.8 true cm}

\pacs{
\hspace{2.5cm}
}

\vskip2pc]
\narrowtext

\vspace{-1 true cm}

\section{Introduction}

Mixed spin chains with alternating spins (1,1/2) are paradigmatic examples
of strongly correlated systems exhibiting  excitations of
different types, namely, both gapless and gapped excitations in the
low energy spectrum \cite{ra,rb,rc,rd}.
This is a manifestation of their ferrimagnetic character and they are
responsible for their unusual properties.
On the contrary,
 one-dimensional integer-spin Anti-ferromagnetic Heisenberg (AFH) model
has a unique disordered ground state with a finite excitation gap while the
half-integer spin AFH chain is gapless \cite{haldane}.
These ferrimagnetic quantum chains have recently attracted much attention
and several techniques have been devoted to their study :
analytically,
such as Spin Wave Theory (SWT) \cite{rb,rc},
Quantum Renormalization Group \cite{ahl}, Variational \cite{ra},
and numerically, such as conventional \cite{rc},
transfer matrix DMRG \cite{rd}, QMC \cite{rd} etc...
Moreover, the interest is increased
due to the fact that experimental realizations
of these one-dimensional ferrimagnets also exist. For instance,
the oxamato compound $Ni Cu (pba) (D_2O)_3 \cdot 2D_2O$
exhibits a structure of a $(S_1,S_2)=(1,1/2)$ ferrimagnetic chain
\cite{oxamato}
where the spins 1 are located at the Ni sites and the spins 1/2 at the
Cu sites. It is found that below $T=7 K$ it exhibits Long Range Order (LRO).
Materials realizing other magnitudes of spins $(S_1,S_2)$
have been also synthesized \cite{others}.
In all these compounds it is assumed that the
interchain coupling is much smaller than the intra-chain interactions so
that the system becomes a set of effectively uncoupled quantum chains.

On the other hand, many interesting investigations have been
devoted to spin ladders. 
They consist of coupled
one-dimensional chains and may be regarded as interpolating one- and
two-dimensional systems. Theoretical studies have suggested two different
universality class for the uniform spin ladders, i.e. the anti-ferromagnetic
spin-${1\over 2}$ ladders are gapful or gapless, depending on whether
$n_l$ (the number of legs) is even or odd \cite{ladders}. Some experimental
observations are in agreement with this results \cite{review}. However
mixed-spin ladders may change this universality. In a recent work \cite{QRG}
we  have considered a ferrimagnetic 2-leg ladder and conjectured its phase
diagram by using an approximate Quantum RG analysis. We have shown that
for the homogeneous antiferromagnetic couplings the model is gapless but
an energy gap may appear in a specific type of configuration of the
coupling constants.

\begin{figure}
\epsfxsize=9cm \epsfysize=7cm  \epsffile{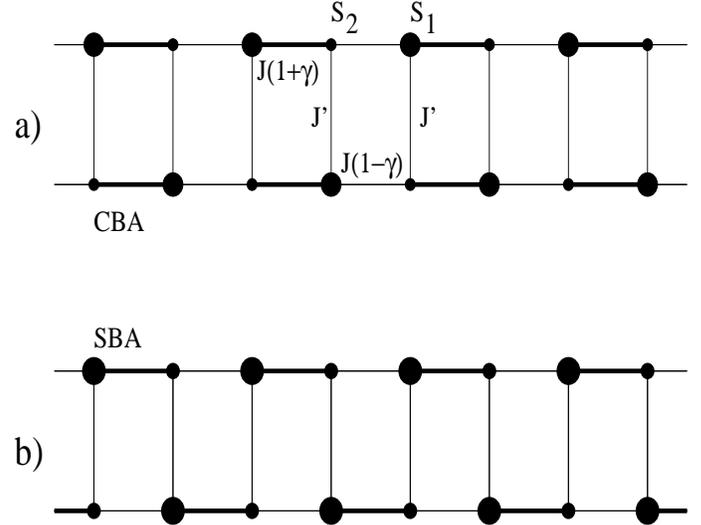}
\narrowtext
\caption[]{ A picture of the dimerization patterns
considered in a 2-leg mixed spin ladder: a) Columnar Bond Alternation (CBA),
b) Staggered Bond Alternation (SBA). Small circles are $S={1 \over 2}$
and large ones are $S=1$.
}
\label{fig0}
\end{figure}
\noindent

In this work we present the results of an extensive numerical analysis
using the Lanczos method  \cite{review} applied to a large class of
ferrimagnetic 2-leg ladders for the first time
(see \cite{ladders} for a review on ladder materials.)
Apart from accurate numerical results which show the effect
of inter-chain interaction in a ferrimagnetic system, we obtain the
phase diagram of Ref.\cite{QRG} with more accuracy.
In the present phase diagram,
which is obtained by Lanczos results,
we specify the phase boundary between the
gapless and gapful phases more accurately and show that the transition between
two gapless phases appears at zero inter-chain coupling and is first-order.
An analytic Spin Wave Theory is also
introduced to describe the low-energy excitations of this model.
Moreover we will consider the effect of bond-alternation on the Spin-Peierls
instability and its conditional or unconditional character.
The model we have considered here, depicted in Fig.1, can be assumed as
the first step to consider the inter-chain interactions in a ferrimagnetic
material.
(For instance, $MnCu(pbaOH)(H_2O)_3$ has such kind of structure \cite{Kahn}.)
In our model  
mixed spins have lengths $S_1=1$ and $S_2=1/2$
and are located at alternating positions
in this bipartite lattice, making up the sub-lattices A and B, respectively.
One may think of two ($1, 1/2$) ferrimagnetic chains which interact via rungs in
a ladder realization
where different kind of spins are located on each rung (see Fig.\ref{fig0}).
Specifically, the isotropic Heisenberg Hamiltonian
we consider is the following ( with $i=2n+\alpha$),

\begin{eqnarray}
H &=& J \sum_{\alpha=1}^2 \sum_{n=0}^{{L\over 2}-1} \; (\;
[1+\gamma^{(\alpha)}(i)]
\; \; {\bf S}^{(\alpha)}_1(i) \cdot  {\bf S}^{(\alpha)}_2(i+1) \nonumber \\ 
&+&[1 + \gamma^{(\alpha)}(i+1)]\;\;
{\bf S}^{(\alpha)}_2(i+1) \cdot  {\bf S}^{(\alpha)}_1(i+2)\; ) \nonumber \\
& + & J^\prime \sum_{n=0}^{{L\over 2}-1}
[{\bf S}^{(1)}_1(2n+1) \cdot {\bf S}^{(2)}_2(2n+1) \nonumber \\
&+&{\bf S}^{(1)}_2(2n+2) \cdot {\bf S}^{(2)}_1(2n+2)]
\;,
\label{1}
\end{eqnarray}

\noindent where ${\bf S}^{(\alpha)}_1(n)$ denotes the quantum spin-$1$
at site $n$ in the leg $\alpha=1,2$ of the ladder, similarly
${\bf S}^{(\alpha)}_2(n)$ for the spin-$1/2$,
$J$ and $J^\prime$ are coupling constants
along the legs and the rungs respectively, and $\gamma^{(\alpha)}(n)$
is the dimerization patterns. We consider two different dimerization
patterns :
i) CBA (Columnar Bond Alternation, Fig.\ref{fig0}(a))
for which $\gamma^{(\alpha)}(n)=(-1)^{n+1}\gamma$ \cite{QRG},
and ii) SBA (Staggered Bond Alternation, Fig.\ref{fig0}(b)) where
$\gamma^{(\alpha)}(n)=(-1)^{\alpha+1+n}\gamma$ \cite{stag1,stag2}.
In order to keep the system always in the antiferromagnetic regime,
we restrict the range of variation as $|\gamma| \leq 1$.
We use periodic boundary conditions along the legs of the ladder.
The number of sites is $N=2\times L$, where $L$ is the length.

The paper is organized as follows: in next section we introduce
the spin wave theory applied to a non-dimerised ($\gamma=0$)
ferrimagnetic ($S_1, S_2$) ladder. Dispersion relations, ground
state energy ,  energy gap and magnetization of this model are
obtained within SWT in order to explain the nature of elementary
excitations in this system. In Sec. III we present our numerical
results obtained with the Lanczos method and we contrast them
against the previous SWT results. In Sec. IV we extend our
Lanczos results to include a numerical analysis of the phase
diagram for a 2-leg mixed spin ladder to specify the boundary
between the gapless and gapful phases. In Sec. V. the
spin-Peierls instability  is discussed by computing the magnetic
energy gain of the present model. We finally present our
conclusions in Sec. VI.

\section{Spin Wave Theory}

In this section we shall perform a mean-field treatment of the
ferrimagnetic ladders and for simplicity
we shall concentrate on the case of zero dimerization $\gamma=0$.

The application of the Lieb-Mattis theorem \cite{lieb-mattis}
to this ferrimagnetic ladder
predicts the existence of a Ground State (GS) with Total Spin $S_G$ given
by $S_G=L(S_1-S_2)$. This leads to the existence of a non-vanishing value
of the magnetization at zero temperature. Then, a natural issue arises,
namely, the study of classical ferrimagnetic order v.s. quantum antiferromagnetic
fluctuations \cite{ahl}. This system presents gapless Ferromagnetic excitations
(Goldstone modes due to broken symmetry) with spin $S_G-1$ and gapped Antiferromagnetic excitations with spin $S_G+1$.

\noindent Thus, as we expect an ordered ground state,
it is natural to  perform a spin wave theory analysis
to describe the low energy quantum fluctuations around a
classical ferrimagnetic ground state.
We shall employ the SWT in the
linear approximation.

To this end, we first make a standard
Holstein-Primakoff transformation on the $(1,1/2)$-spins
of eq. (\ref{1}).
Let $a^{(\alpha)}_k$ be the modes in momentum space associated
to the sub-lattice A for the legs $\alpha=1,2$, and $b^{(\alpha)}_k$ similarly for
the modes in the sub-lattice B.
The Hamiltonian can be written in the following form.
$$
H_{SWT}=H_{rung} + \sum_{\alpha=1,2} H_{leg}^{(\alpha)}
$$
\begin{eqnarray}
&H&_{rung}=-LJ'S_1S_2 \nonumber \\
&+&J'\sum_k [ S_1(a_k^{\dagger(1)} a_k^{(1)}
+a_k^{\dagger(2)} a_k^{(2)})
+ S_2(b_k^{\dagger(1)} b_k^{(1)}
+b_k^{\dagger(2)} b_k^{(2)}) ] \nonumber \\
&+&J'\sqrt{S_1S_2} \sum_k (a_k^{(1)}b_k^{(2)}+a_k^{(2)}b_k^{(1)}
+a_k^{\dagger (1)}b_k^{\dagger(2)}+a_k^{\dagger (2)}b_k^{\dagger(1)}) \nonumber
\end{eqnarray}
\begin{eqnarray}
H_{leg}^{(\alpha)}&=&-LJS_1S_2+2J\sum_k (S_1 a_k^{\dagger (\alpha)} a_k^{(\alpha)}
+S_1 b_k^{\dagger (\alpha)} b_k^{(\alpha)}) \nonumber \\
&+&J\sqrt{S_1S_2} \sum_k
2Cos(k) (a_k^{(\alpha)} b_k^{(\alpha)} + a_k^{\dagger (\alpha)}
b_k^{\dagger (\alpha)})
\label{swt1}
\end{eqnarray}

\noindent As it is apparent, this is not enough to diagonalize
the Hamiltonian with a Bogoliubov transformation because the
degrees of freedom of each leg appear coupled together.
Fortunately enough, we can devise a second transformation by introducing
a couple of symmetric $s',s''$ and antisymmetric $a',a''$ fields as follows:

\begin{eqnarray}
a^{(\alpha)} = {1\over \sqrt{2}} (s'_k + (-1)^{1+\alpha} a'_k) \;, \nonumber \\
b^{(\alpha)} = {1\over \sqrt{2}} (s''_k + (-1)^{1+\alpha} a''_k) \;,
\label{2}
\end{eqnarray}

 \noindent with $\alpha=1,2$ represents leg index. Then, it is possible to show that
the resulting SWT Hamiltonian takes the following form in terms of these fields:

\begin{equation}
H_{SWT}=-LS_1S_2(2J+J')+h(s)+h(a) \;,
\label{3}
\end{equation}

where

\begin{eqnarray}
h(s)&=&C_1\sum_k s'^{\d}_k s'_k +C_2\sum_k s''^{\d}_k s''_k \\
& + &\sqrt{S_1S_2} \sum_k (J 2\cos k +J')(s'_k s''_k+s'^{\d}_k s''^{\d}_k) \;,
\nonumber \\
h(a)&=&C_1\sum_k a'^{\d}_k a'_k +C_2\sum_k a''^{\d}_k a''_k \\
& + &\sqrt{S_1S_2} \sum_k (J 2\cos k -J')(a'_k a''_k+a'^{\d}_k a''^{\d}_k) \;,
\nonumber
\label{4b}
\end{eqnarray}

\noindent and now using a set of Bogoliubov transformations we will arrive at
the following expressions for these Hamiltonians $h(s)$, $h(a)$ as,

\begin{eqnarray}
h(s)&=&\sum_k [\zeta^{(+)}(k)+\omega_A^{(+)}(k) A_k^{\d (+)}A_k^{(+)}
+(A_k^{(+)} \leftrightarrow B_k^{(+)})] \;,
\nonumber \\
h(a)&=&\sum_k [\zeta^{(-)}(k)+
\omega_B^{(-)}(k) B_k^{\d (-)}B_k^{(-)} + (B_k^{(-)} \leftrightarrow A_k^{(-)})] \;,
\nonumber \\
  \;\;
\label{6}
\end{eqnarray}
\noindent
\begin{figure}
\epsfxsize=9cm \epsfysize=7cm  \epsffile{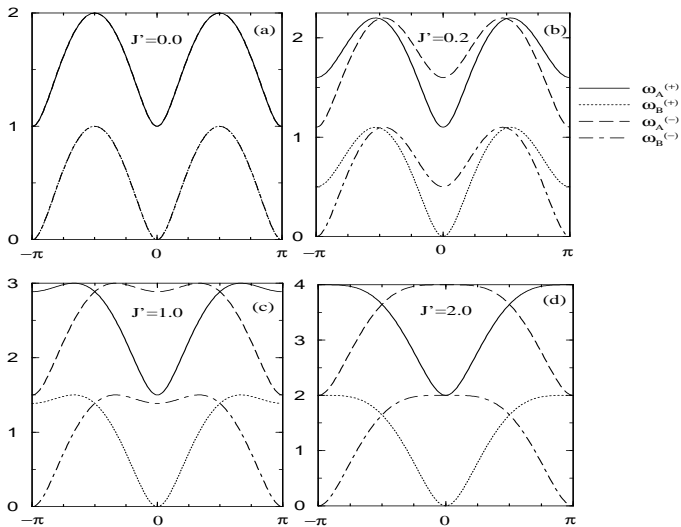}
\narrowtext
\caption[]{ Dispersion relations for a ferrimagnetic ladder
for different values of rung coupling namely $J'=0.0, 0.2, 1.0, 2.0$.
The $J'=0$ case (a) is the same as one-dimensional ferrimagnets.
}
\label{fig1}
\end{figure}
\noindent

\noindent where $A_k^{\d (\pm)},B_k^{\d (\pm)}$ are the quasi-particle creation operators and
the frequencies $\omega_A^{(\pm)}(k), \omega_B^{(\pm)}(k)$
of the several excitations take the following expressions as a function of
the model coupling constants,

\begin{eqnarray}
\omega_A^{(\pm)}(k)&=&\frac{1}{2}(C_1-C_2+\sqrt{(C_1+C_2)^2-4C_3^{2 (\pm)}(k)}) \;,
\nonumber \\
\omega_B^{(\pm)}(k)&=&\frac{1}{2}(C_2-C_1+\sqrt{(C_1+C_2)^2-4C_3^{2 (\pm)}(k)}) \;,
\nonumber \\
\zeta^{(\pm)}(k)&=&-\frac{1}{2}(C_1-C_2)+\frac{1}{2}
\sqrt{(C_1+C_2)^2-4C_3^{2 (\pm)}(k)} \;,
\nonumber \\
C_1&=&(2J+J')S_1 \hspace {1cm} C_2=(2J+J')S_2  \;, \nonumber \\
&C&_3^{(\pm)}(k)=\sqrt{S_1S_2}(2J \cos k \pm J') \;,
\label{7}
\end{eqnarray}

\noindent

\begin{figure}
\hspace{-0.8cm}
\epsfxsize=9cm  \epsfysize=7cm \epsffile{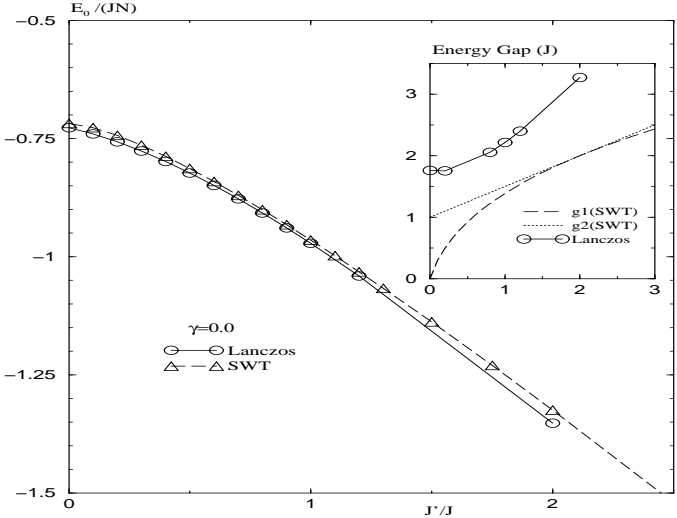}
\narrowtext
\caption[]{Lanczos v.s. SWT results for GS energy,
as a function of the interchain coupling $J'$. Inset similarly shows
the Lanczos results v.s. SWT for energy gaps.
}
\label{fig2}
\end{figure}

In Fig.\ref{fig1} we have plotted the four branches of the
dispersion relations $\omega_A^{(\pm)}(k), \omega_B^{(\pm)}(k)$.
For vanishing interchain coupling $J'=0$ (Fig. \ref{fig1} (a)) we
reproduce the 1d-chain result. As the interaction is switched on,
we appreciate the splitting of each zero-coupling branch into
another two new branches in the case of ladder. The appearance of
four modes is related to defining two boson operators for each
leg of ladder. At $k=0$ we define the gap as $g_1\equiv
|\omega_B^{(+)}(0)-\omega_B^{(-)}(0)|$, which goes to zero with
vanishing $J'$ as can be seen in Fig.\ref{fig2}. We also define
the gap $g_2\equiv |\omega_A^{(+)}(0)-\omega_B^{(+)}(0)|$, which
remains finite $\forall J'$  (see Fig.\ref{fig2}). So we consider
$g_2$ as the actual energy gap which is defined by SWT. The
lowest energy dispersion relation is $\omega_B^{(+)}$ which shows
gapless excitations at $k=0$,
\begin{equation}
\omega_B^{(+)}(k\rightarrow 0)=\frac{2S_1S_2 J}{|S_1-S_2|}k^2
\label{swt2}
\end{equation}
The first excited branch which shows an actual gap for any value of $J'$ is
$\omega_A^{(+)}$ with the following behaviour in the long wave length limit,
\begin{equation}
\omega_A^{(+)}(k\rightarrow 0)=(2J+J')|S_1-S_2| +\frac{2S_1S_2 J}{|S_1-S_2|}k^2
\label{swt3}
\end{equation}

Eq.(\ref{swt2}) is similar to a ferromagnetic dispersion relation, then we
expect a ferromagnetic behaviour for these excitations. Moreover it can be
shown that the $\omega_B^{(+)}(k)$ modes are created by acting on the classical
ferrimagnetic ordered state ($|0\rangle$) with
$S_k^-=\frac{1}{\sqrt{L}}\sum_n\sum_{\alpha}(S_1^{-(\alpha)}(n)+
S_2^{-(\alpha)}(n))e^{ikn}$, where $k\neq0$. The outcomes of this operation
are those states which have $S_{tot}^z=L|S_1-S_2|-1$. These states are
annihilated by $S_{tot}^+$. Therefore we conclude that these states have
$S_{tot}=L|S_1-S_2|-1$ which is another sign for the ferromagnetic behaviour
of  $\omega_B^{(+)}$ modes. The magnetization of $S_1=1$ sub-lattice shown in
Fig.(\ref{fig4}) verifies the reduction of magnetization from its classical
value ($\langle S_1^z\rangle_{class.}=1$) which shows that the quantum
fluctuations of the lowest branch has a ferromagnetic property. Similarly,
the $\omega_A^{(+)}$ modes are created by acting on $|0\rangle$ with
$S_k^+=\frac{1}{\sqrt{L}}\sum_n\sum_{\alpha}(S_1^{+(\alpha)}(n)+
S_2^{+(\alpha)}(n))e^{ikn}$, where $k\neq0$.
It is also possible to show that those states have $S_{tot}=L|S_1-S_2|+1$
which is the property of an anti-ferromagnetic excitations, i.e. the
excited states have larger spin than the ground state.
All of these explanations for the excited states are examined for a
finite system size ($N=2L=8, 12, 16, 20$) by Lanczos method where we will
discuss  them in the next section.

Using the quantities
$\zeta^{(\pm)}(k)$ and eqs. (\ref{2})-(\ref{7}), we have also computed the
GS energy per site $E_0/JN$ as follows,

\begin{equation}
E_0=-NS_1S_2(2J+J')+\sum_k (\zeta^{(+)}(k) + \zeta^{(-)}(k)).
\label{7b}
\end{equation}
It is plotted
as a function of $J'$ in Fig.\ref{fig2}.

\noindent

In a similar way the sub-lattice
magnetization $m_1$ as a function of the
couplings is calculated,
\begin{equation}
m_1=\frac{1}{L} \sum_{i=1}^{L/2} \langle  S^{z(1)}_1(i) + S^{z(2)}_1(i) \rangle.
\label{7c}
\end{equation}
The other sub-lattice magnetization
$m_2=\langle S^z_2 \rangle$ can also be obtained by using the identity :
$m_1+m_2=S_1-S_2$. We have plotted $m_1$ as a function of $J'$ in
Fig.\ref{fig4}. We will discuss on this result in the next section.

\section{Numerical Analysis: Lanczos Method}

Let us now present our numerical Lanczos results.

\noindent We have studied ladders with a number of sites
$N=2\times L$ with $L=4,6,8,10$ due to the constraint of
periodic BC's. We always set $J=1$ and vary $J'=0.0,0.2,0.8,1.0,1.2,1.8,2.0$. The dimerization parameter
is varied from $\gamma=0.0$ to $\gamma=1.0$ in steps of $0.2$
for both CBA and SBA patterns.

\noindent As a testground, we have checked several quantities at $J'=0$, like
the GS energy  per site $e_0$ and Antiferromagnetic gap $\Delta_0$
and we find excellent agreement with
the DMRG results for the spin chain as shown in Table 1, namely,
$e_0(DMRG)=-0.727$  and $\Delta_0(DMRG)=1.759$ \cite{rc,rd}.

\noindent In Fig. \ref{fig2} we plot the variation with $J'$ of the GS energy and we
find that the extrapolated numerical results agree very well with the SWT results
in the whole range.
However, for the Antiferromagnetic gap (see inset of Fig.\ref{fig2}) we
find clear numerical differences with SWT. For comparison, we have plotted the
two gaps, $g_1,g_2$ introduced  before, because for the strong coupling region
around
$J' \sim 2$, both gaps turn out to be very close to each other.
We clearly find that the exact gap obtained with Lanczos is well above the
values obtained with the SWT approximation.

\begin{figure}
\vspace{-0.35cm}
\hspace{-0.8cm}
\epsfxsize=9cm \epsfysize=8cm  \epsffile{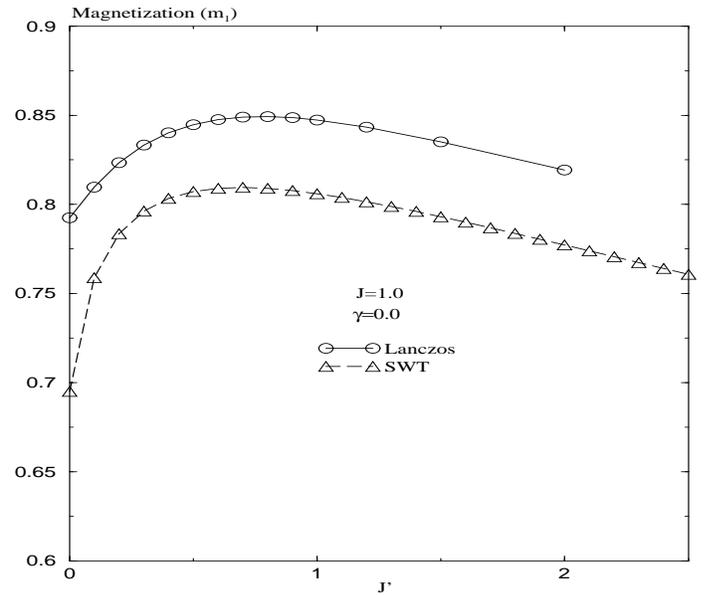}
\narrowtext
\caption[]{Sub-lattice magnetization of spin $S_1=1$ ($\langle S_1^z \rangle$),
Lanczos v.s. SWT results as a function of $J'$.
}
\label{fig4}
\end{figure}
\noindent

Likewise, we have computed several
sub-lattice magnetization and in Fig. \ref{fig4} we show
$m_1 = \langle {\bf S}^z_1(1) \rangle$ for the spin-1 case at site $n=1$ and
we have checked that it is the same at all other positions due to translational
invariance. We find that the SWT calculation essentially underestimates the
Lanczos result.
The discrepancies between  SWT results for energy gap
(inset of Fig. \ref{fig2}) and magnetization (Fig. \ref{fig4}) means  that the
linear SWT does not provide us with a complete account of
quantum fluctuations around the classical ferrimagnetic order.
However the ground state energy obtained
by both methods are in good agreement. For instance in the homogeneous case
($\gamma=0, J=1, J'=1$)  the SWT result is $e_0(SWT)=\frac{E_0}{JN}=
-0.96475$ which has only $0.7$ percent error with respect to
Lanczos result (Table.1).

\noindent
\begin{figure}
\vspace{-0.15cm}
\hspace{-0.8cm}
\epsfxsize=9cm \epsfysize=8cm  \epsffile{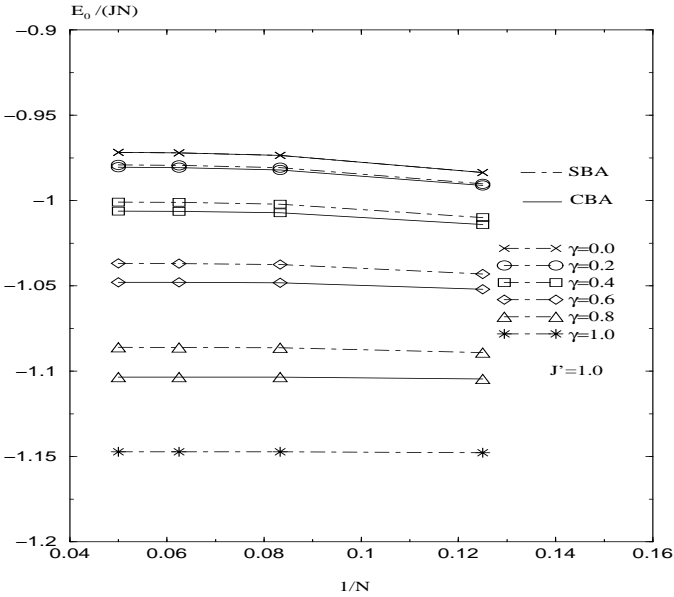}
\narrowtext
\caption[]{Plot of the GS energy (Lanczos) v.s. $1/N$
for CBA and SBA dimerization patterns. Convergence to thermodynamic
limit is very sharp for all dimerization value ($\gamma$).
}
\label{fig5}
\end{figure}
\noindent

\noindent We have also  studied the effect of dimerization in the ferrimagnetic
2-leg ladder in the Columnar (CBA) and staggered (SBA) configurations.
In Fig.\ref{fig5} we present the results of Lanczos calculations for $J'=1$ and
varying the staggering parameter $\gamma$ in the whole range.
We first check that our results are well converged for the sizes considered
(N=8,12,16,20).
The ground state energy converges to the value of thermodynamic limit
as rapid as
in the one-dimensional ferrimagnets. It means that the finite size effects
are small in the ferrimagnetic ladders as well as in 1-D models.
We then observe that the effect of dimerization ($\gamma$) is generically
to decrease the GS energy and that the columnar configuration
produces a bigger energy decrease than the staggered pattern.
This  shows that the CBA pattern is more stable than the SBA
configuration. The stability is improved by increasing dimerization.

\begin{center}
\begin{tabular}{|c|c|c|c|c|c|c|}
\hline
$J'$   & 0.0   & 0.1 & 0.2 & 0.3 & 0.4 & 0.5  \\
\hline
$e_0$  & $-0.7270$ & $-0.7403$ & $-0.7569$ & $-0.7764$ & $ -0.7985$ & $-0.8229$ \\
\hline
$J'$   & 0.6   & 0.7 & 0.8 & 1.0 & 1.2 & 2.0  \\
\hline
$e_0$  & $-0.8493$  & $-0.8776$ & $-0.9075$ & $-0.9716$ & $-1.0408$ & $-1.3524$   \\
\hline
$J'$   & 0.0   & 0.2 & 0.8 & 1.0 & 1.2 & 2.0  \\
\hline
$\Delta_0$  & $ 1.7589$ & $ 1.7544$ & $2.0545$ & $2.2158$ & $2.3981$ &
$3.2696$ \\
\hline
\end{tabular}
\end{center}
\begin{center}
Table 1.
Extrapolated ($N\rightarrow \infty$) Lanczos results
for GS density energy $e_0=E_0/JN$ and
AF-gap
$\Delta_0$.
\end{center}

\begin{figure}
\hspace{-0.8cm}
\epsfxsize=9cm \epsfysize=8cm  \epsffile{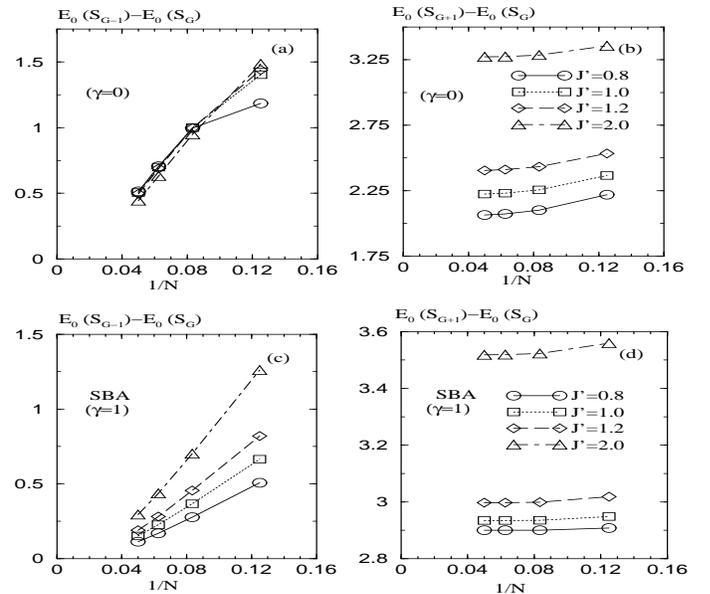}
\narrowtext
\caption[]{Plot of energy gaps (Lanczos): (a)-(b) Ferro-gap and AF-gap
without dimerization v.s. $1/N$ for several couplings $J'$.
(c)-(d) Same with dimerization (SBA).
}
\label{fig6}
\end{figure}

The Lanczos analysis of the ferromagnetic and antiferromagnetic
gaps are shown in Figs.\ref{fig6}. In Fig.\ref{fig6}(a) we clearly find that in the absence of
dimerization $\gamma=0$, the Ferro-gap $E_0(S_{G}-1)-E_0(S_{G})$
scales to zero as the system size $N$ goes to $\infty$.
This corresponds to the gapless excitations in the SWT analysis.
We have plotted data for different values of $J' (0.8, 1.0, 1.2, 2.0)$
which shows the same behaviour. We have also checked for some more values of
$J'$ and the same results are obtained.
Moreover, we have found that this also holds true when dimerization is
present and in Fig.\ref{fig6}(c) we show this for the SBA configuration
which shows that we have  gapless excitations with ferromagnetic nature
for all values of $J'$ and the dimerization parameters ($\gamma$).
In Figs.\ref{fig6}(b),(d) we have plotted  the AF-gap $E_0(S_{G}+1)-E_0(S_{G})$
for some values of $J'$ and $\gamma$.
We find that
the AF-gap remains finite in the thermodynamic limit, moreover
the interchain coupling $J'$ increases the gap both with or without
dimerization.
The notion of Ferro-gap and AF-gap is related to the excitations with spin
lower or higher than the spin of the ground state, respectively.
This is in complete agreement with the SWT explanation of the low energy
spectrum.

We have presented the extrapolated values of ground state energy per site
($e_0$) and AF-gap ($\Delta_0$) in Table 1. 
The scaling form which we 
have considered for the extrapolation to $N\rightarrow\infty$ is a power-law
fit. We have implemented the following function for the scaling form of the
ground state energy,
\begin{equation}
e_0(N)=e_0(\infty)+\frac{a}{N^{\nu}}
\label{extrapolation}
\end{equation}
\noindent where $a$ and $\nu$ are constants and obtained to have least-square
root error. We have examined a polynomial form but the best fitting was obtained 
by the power-law fit of Eq.(\ref{extrapolation}). In the case of energy gap
we considered both Eq.(\ref{extrapolation}) and an exponential form, 
(i.e. $\Delta_0(N)=\Delta_0(\infty)+B \;exp(-|c|N)/(N^{\mu})$) where 
$B, c$ and $\mu$ are constants. Both scaling forms give us the same results for 
$\Delta_0(\infty)$ and 
fit very well to the data of Fig.(\ref{fig6}).
Although from the SWT calculations we obtained the $k^2$ dependence for both
gapless and gapped spectrum (Eqs.(\ref{swt2}, \ref{swt3})) which impose that the scaling form should be
like $N^{-2}$, however this type of function does not fit to over data and leads to
a big value for the square least error.

\begin{figure}
\hspace{-0.8cm}
\epsfxsize=9cm \epsfysize=8cm  \epsffile{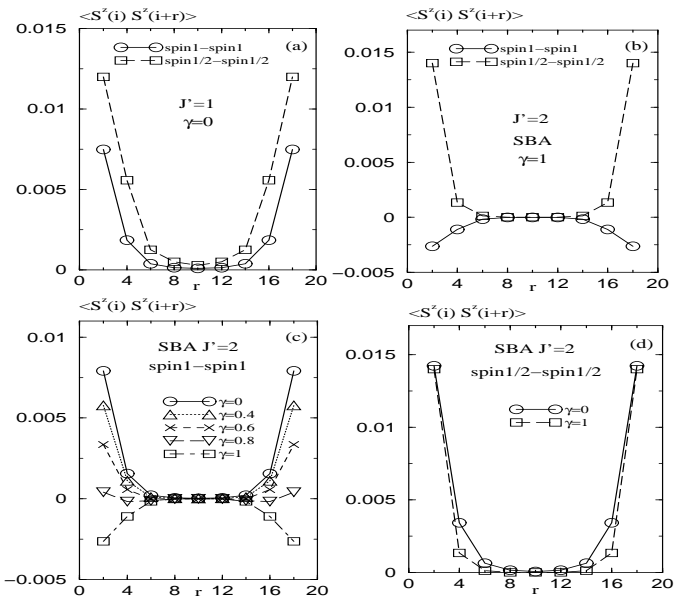}
\narrowtext
\caption[]{Correlation functions for both spin (1/2, 1/2) and
spin (1, 1) (Lanczos):
(a) without dimerization; (b)-(d) with dimerization (SBA).
}
\label{fig7}
\end{figure}

In Figs.\ref{fig7} we present our  numerical analysis of the two-point GS correlation
functions (Eq.(\ref{correlation}))
for spin-1-spin-1 $(\sigma,\sigma')=(1,1)$
and spin-1/2-spin-1/2 $(\sigma,\sigma')=(1/2,1/2)$.
In Fig.\ref{fig7}(a) we observe that without dimerization $\gamma=0$,
the spin-1/2-spin-1/2 correlation has a slower exponential fall-off than the
spin-1-spin-1 case
which means that the quantum fluctuations for spin-$1/2$ are
stronger   
than the spin-$1$ case. Moreover, they both are positive which
means that the spins are aligned parallels in each sub-lattice
(ferro-magnetically). In Fig.\ref{fig7}(b) we make a check by
comparing with the one-dimensional case. This is achieved by using
the ``snake mechanism'' \cite{stag1}: we set $\gamma=1$ and
$J'=2$ in eq. (\ref{1}) for the SBA configuration. Under these
conditions, we are dealing with an ordinary ferrimagnetic chain
with alternating spins of magnitude 1/2 and 1 without dimerization
and with effective coupling constant $J_{eff}=2.$ Thus, as we are
plotting the connected correlation function

\begin{equation}
\langle S^z_0 S^z_n\rangle - \langle S^z_0\rangle \langle S^z_n\rangle
\label{correlation}
\end{equation}

\noindent then
the case spin 1-1 has opposite sign and happens only in
one-dimensional limit ($J'=2, \gamma=1, SBA$). It is due to
strong quantum fluctuations in the one-dimensional case.
This is in full agreement with results obtained earlier in \cite{rc}
(see Fig. 5(a) and (b)  of this reference.)
We have reproduced their results in a particular case of our study.
Moreover, we have extended this numerical analysis for real 2-leg
mixed spin ladders as explained in the other figures.

In order to see what is the effect of the interchain coupling $J'$ on the
correlation length $\xi$ in the 2-leg ladder as compared with the pure
one-dimensional case, we plot in Fig.\ref{fig7}(d) the spin-1/2-spin-1/2 case
for $\gamma=0$ and $\gamma=1$ at $J'=2$ and we observe that in the ladder
($\gamma=0$) there is a slower decay than in the chain ($\gamma=1$).
This signals that the correlation length is bigger  in the ferrimagnetic ladder
than in the corresponding chain: $\xi_{ladder} > \xi_{chain}$.
Likewise, in Fig. \ref{fig7}(c)  we have made a similar analysis to see
the effect of the
dimerization in the staggered configurations (SBA).
We observe that in the ladder
$\gamma \neq 1$    
the correlation length $\xi$ decreases as the dimerization
gets stronger. Also, the correlations remain ferromagnetic in
this spin-1 sub-lattice.

\section{Numerical Phase Diagram}

With the numerical tools provided by the Lanczos method we
can also make an analysis of the phase diagram exhibited by
the 2-leg mixed spin ladders in the space of couplings of
the dimerization $\gamma$ and the interchain coupling constant $J'/J$.
We have performed this analysis for both antiferromagnetic couplings
$J'>0$ and ferromagnetic couplings $J'<0$ (in units of $J=1$).
Moreover, we have studied both dimerization patterns, CBA and SBA,
considered throughout  this work.

In order to set up this phase diagram we determine numerically the
existence or lack of an energy gap in the lowest lying spectrum,
as we have done in previous sections. For antiferromagnetic $J'$
we find gapless ferrimagnetic order for both CBA and SBA
configurations. This ferrimagnetic order is supported by a
non-zero value of magnetization. This is in agreement with the
approximate QRG analysis in \cite{QRG}, and thus we do not
explicitly plot this diagram. The most interesting
characteristics are found in the ferromagnetic $J'$ region. In
this case, we find that the GS is a spin singlet while the lowest
excitation is a spin triplet. For the SBA dimerization pattern we
always find a gapless behaviour for any value of the dimerization
parameter $\gamma$. Since the ground state is a singlet the
magnetization is zero for all $J'<0$ region. Thus we conclude
that the phase boundary between the two gapless phases ($J'<0$
and $J'>0$) is at $J'=0$ where a first-order phase transition
occurs. If we consider the magnetization as the order parameter
then it is discontinuous at $J'=0$ (1-st order transition). The
$J'=0$ transition line modifies the location of phase boundary
obtained by QRG in the previous study (see Fig.3 in
Ref.\cite{QRG}).

\begin{figure}
\hspace{-0.8cm}
\epsfxsize=9cm \epsfysize=8cm  \epsffile{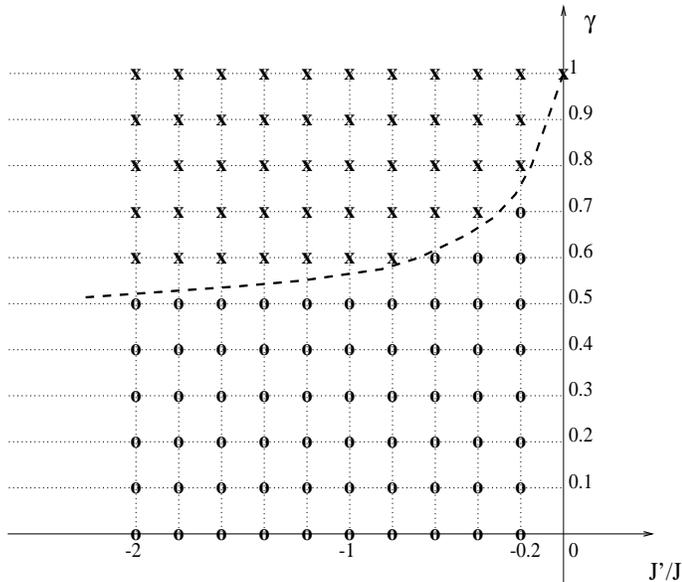}
\narrowtext
\caption[]{Phase diagram of a 2-leg mixed spin ladder
with CBA dimerization in the space of couplings
$\gamma$ v.s. $J'/J$ ($J'<0$). A cross $X$ symbol denotes a finite gap
while an $O$ symbol denotes a zero gap. The dashed line is an estimate for
the phase boundary between the gapless and gapful phases.
}
\label{fig9}
\end{figure}

However, the situation turns  richer for the CBA dimerization
and this is the case that we plot in Fig.\ref{fig9}.

\noindent To obtain this numerical phase diagram we have split
the range $\gamma \in [0, 1]$ into $0.1$ steps and the range
$J' \in [-2, 0)$ into $0.2$ steps. Thus, we end up with a
$10\times 10$ grid of numerical points.
For each point in this grid, we have computed the lowest energy
state in the sectors of total spin component $S^z=0,1$.
This is done for a series of lattice sizes, namely,
$N=2\times L$ with $L=4,6,8,10$. With this data we extrapolate
the value of the energy gap for infinite lattices and draw
a symbols $X$ if the gap is non-vanishing and a symbol $O$
otherwise. These extensive Lanczos analysis is presented in
Fig.\ref{fig9}.

\noindent Here we find two phases depending on whether we are in a strong
or weak dimerization regime, assuming also that the interchain
coupling $J'$ is not negligible.
We clearly detect a gapful phase in the upper part of the diagram
and a gapless phase in the lower part. From our numerical data we also
estimate the phase separating line
plotted as a dashed line in Fig.\ref{fig9} and we observe that it
approaches a horizontal asymptote as the interchain coupling
becomes very large.
As it has been mentioned before the whole $J'<0$ region has zero magnetization
due to the spin of ground state which is $S=0$. So the phase boundary between
the two gapless phases ($J'<0$ and $J'>0$) is at $J'=0$ accompanying a
first-order transition similar to SBA configuration.


\section{The Spin-Peierls Instability}

We have also carried out a Lanczos study of the spin-Peierls (SP)
instability. This  phenomena has been  realized in several
materials and here we anticipate this possibility in the
ferrimagnetic 2-leg ladders. This transition towards a dimerised
GS ferrimagnetic ladder is determined by the competition between
the lowering of the magnetic energy due to dimerization and the
raising of the elastic energy due to phonons. According to
standard terminology, the transition is called unconditional if
the ground state is dimerised for arbitrary large value of the
spin-phonon coupling (or small value of dimerization), and it is
called conditional otherwise.

Using exact diagonalization techniques it has been possible to stablish
\cite{s-P} that integer spin chains do not have a spin-Peierls transition
while half-integer spin chains does exhibit this transition.
This behaviour was predicted by Schulz \cite{schulz} using bosonization
techniques. However, other analytical studies had contradicted this
conclusion \cite{s-P}.
Therefore, the study of the spin-Peierls transition provides us with an
alternative view of the quantum differences between integer and half-integers
spins, in addition to the more familiar gapped v.s. gapless behaviour.

Moreover, it is very interesting to study the SP transition
for {\it quasi}-1D systems as a function of the interchain coupling $J'$.
This is usually done within the so called chain mean-field theory which
underestimates the effect of this coupling. Here we can treat its effect without any bias.

\begin{figure}
\hspace{-0.8cm}
\epsfxsize=9cm \epsfysize=8cm  \epsffile{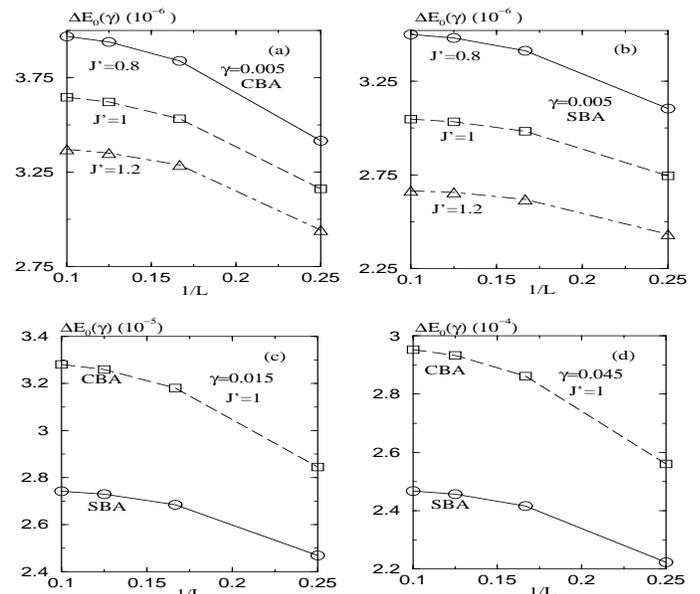}
\narrowtext
\caption[]{Plot of the magnetic energy gain (Lanczos)
$\Delta E_0(\gamma)$ eq. (\ref{10}) v.s. $1/L$ for both CBA and SBA
configurations (spin-Peierls analysis).
(a) and (b) for different values of inter-chain coupling ($J'$) and fixed value
of $\gamma=0.005$. Fixed value of $J'=1$ is presented in (c) $\gamma=0.015$ and
(d) $\gamma=0.045$.
}
\label{fig8}
\end{figure}
\noindent

The magnetic energy gain per bond $\Delta E_0(\gamma)$ is computed as follows,

\begin{equation}
\Delta E_0(\gamma) = [E_L(0) - E_L(\gamma)]/L_b ,
\label{10}
\end{equation}

\noindent where $L_b=3L$ is the number of interacting bonds. It
can be shown that studying the limit of (\ref{10}) as
$L\rightarrow \infty$ for small $\gamma$ reveals the conditional
or unconditional character of the SP transition. Specifically,
this feature is encoded in the concavity or convexity character,
respectively, of the curve of (\ref{10}) v.s. $1/L$ \cite{s-P}.
Namely, for a finite system the condition for having an
unconditional dimerised transition is that in the limit of small
$\gamma$, the energy gain (\ref{10}) must strictly increase  as
$N$ increases. This means that the stabilization energy always
overcomes the elastic energy. Otherwise, if the magnetic energy
gain does not strictly increase with increasing the size, then
the transition is conditional,i.e., it depends on the stiffness
of the lattice. This characterization of the conditional v.s.
unconditional character of the dimerization transition can be
recasted in terms of the geometry of the curve for the energy
magnetic gain (\ref{10}) as a function of the size $1/L$: when
the transition is unconditional the curvature of this function is
oriented upwards with respect to the $x$ axis and we call this a
convex curve (see Fig.1(a)) in \cite{s-P}); while in case of the
conditional transition, the curvature is oriented downwards with
respect to the horizontal axis and we call this a concave curve
(see Fig.1(b) in \cite{s-P}.) In Figs. \ref{fig8} we present our
results. We observe that these curves are clearly concave for any
values of the $J'$ couplings that we have investigated. Moreover,
this result holds true for both CBA and SBA configurations. Thus,
we find that the SP transition is conditional for a set of values
in $J'$ ranging from $0$ to $2.0$. We have presented a reduced
number of plots for the variation of $J'$ in Fig.\ref{fig8}
because for larger values the curves have a different scale and
cannot be fitted into one plot.

We have checked this by taking three different value of $\gamma$ ($0.005, \;
0.015, \;0.045$). It has been also shown that the magnetic energy gain in CBA
is bigger than in the SBA configuration
which is in agreement with Fig.(\ref{fig5}) where the CBA configuration is
more stable than the SBA one against small perturbations.

\section{Conclusions}

This paper represents the first extensive numerical study
using the Lanczos method of quantum ferrimagnetism in
quasi-one dimensional models with a ladder structure.
Previous numerical works have dwell upon truly one dimensional
spin chains with alternating spins of magnitude $(S_1,S_2)=(1,1/2)$.

We have presented our results regarding ground state properties
of a 2-leg ladder with two patterns of dimerization: Columnar (CBA)
and Staggered(SBA). Likewise, we have seen how the GS properties
evolve upon variation of the interchain coupling constant $J'$ of
the ladder.

Among the properties we have computed we mention the ground state energy,
Antiferromagnetic gap, gapless behaviour of the Ferromagnetic excitations
as we extrapolate our results to the thermodynamic limit, GS correlation
functions and so on and so forth.
These results have been contrasted with an approximate Spin Wave Theory
analysis in the linear approximation and found a qualitative good agreement
in some properties like the ground state energy, but not for the gaps.
Moreover, we have performed a numerical analysis of when a spin-Peierls
transition can occur and we have reported that this transition is conditional.

We have completed our Lanczos study of 2-leg mixed spin ladders with the
inclusion of a numerical phase diagram which allows us to clarify the different
gapless or gapful phases in the space  of couplings of the model.

We believe that these results are of interest for
researchers in the area of quantum spin systems that may want to know how
the  ferrimagnetism of a  one-dimensional spin chain evolves when an extra chain
is introduced and treated on equal footing.


{\bf Acknowledgements}
We acknowledge discussions with
M. Abolfath,  H. Hamidian, I. Peschel  and G. Sierra.
We would thank the Centro de Supercomputaci\'on Complutense
for the allocation
of CPU time in the SG-Origin 2000 Parallel Computer and also
Max-Planck-Institut f\"ur Physik komplexer Systeme for
computer time.
M.A.M.-D. was supported by the DGES spanish grant
PB97-1190.

\vspace{-0.5 true cm}


\begin{thebibliography}{99}
\vspace{-1.5 true cm}

\bibitem{ra} A.~K.~Kolezhuk, H.-J.~Mikeska, and S.~Yamamoto, Phys. Rev.
{\bf B55}, R3336 (1997).
\bibitem{rb}S. Brehmer, H.-J. Mikeska and S. Yamamoto, J. Phys. Cond. Matt. {\bf 9},
3921 (1997);
\bibitem{rc} S.~K.~Pati,
S.~Ramasesha, and D.~Sen,
Phys. Rev. B {\bf 55}, 8894 (1997).
J. Phys. Cond. Matt. {\bf 9}, 8707 (1997);

\bibitem{rd} S. Yamamoto,
S. Brehmer and H.-J. Mikeska
 Phys. Rev. {\bf B57}, 13610 (1998);
S. Yamamoto and T. Fukui, Phys. Rev. {\bf B57}, R14008 (1998);
S. Yamamoto, T. Fukui, K. Maisinger and U. Schollwock,
J. Phys. Cond. Matt. {\bf 10}, 11033 (1998).

\bibitem{haldane} F. D. M. Haldane, Phys. Rev. Lett. {\bf 50}, 1153 (1983);
Phys. Lett. {\bf A93 }, 464 (1983).

\bibitem{ahl} M. Abolfath, H. Hamidian and A. Langari, cond-mat/9901063.

\bibitem{oxamato} Hagiwara et al.,
J. Phys. Soc. Jpn. {\bf 67}, 2209 (1998).
\bibitem{others} Verdager et al.
Phys. Rev. {\bf B29}, 5144 (1984);
Pei et al.,
Inorg. Chem. {\bf 27}, 47 (1988).


\bibitem{ladders} E. Dagotto and T. M. Rice, Science {\bf 271}, 618
(1996).

\bibitem{review} E. Dagotto, Rev. Mod. Phys. {\bf 66}, 763 (1994).


\bibitem{QRG}
A.~Langari, M.~Abolfath and  M.A. Martin-Delgado,
Phys. Rev. {\bf B61}, 343 (2000).

\bibitem{Kahn} O. Kahn, Y. Pei, M. Verdaguer, J.P. Renard and
J. Sletten, J. Am. Chem. Soc. {\bf 110}, 782 (1988).


\bibitem{stag1} M. A. Martin-Delgado, R. Shankar and G. Sierra, Phys. Rev. Lett.
{\bf 77}, 3443 (1996).
\bibitem{stag2}M. A. Martin-Delgado, J. Dukelsky
and G. Sierra, Phys. Lett. {\bf A 250}, 431 (1998).

\bibitem{lieb-mattis} E. Lieb and D. Mattis, J. Math. Phys. {\bf 3}, 749 (1962).


\bibitem{s-P} D. Guo, T. Kennedy and S. Mazumdar ,Phys. Rev.
{\bf B41}, R9592 (1990).

\bibitem{schulz} H.J. Schulz, Phys. Rev.
{\bf B34}, 6372 (1986).





\end{thebibliography}
\end{document}